\documentclass[9pt,twocolumn,twoside]{pnas-new}

\templatetype{pnasresearcharticle} 

\usepackage{graphicx}
\usepackage{dcolumn}
\usepackage{bm}
\usepackage{amsmath}
\usepackage{xcolor}

\newcommand{\hrho}{\hat{\rho}}
\newcommand{\MC}{\mathcal}
\newcommand{\BF}{\mathbf}

\title{Emergent structural correlations in dense liquids}

\author[a,1]{Ilian Pihlajamaa}
\author[a,1]{Corentin C. L. Laudicina} 
\author[a,b]{Chengjie Luo} 
\author[a,2]{Liesbeth M. C. Janssen}

\affil[a]{Soft Matter \& Biological Physics, Department of Applied Physics,
Eindhoven University of Technology, P.O. Box 513, 5600MB Eindhoven, The Netherlands}
\affil[b]{Max Planck Institute for Dynamics and Self-Organization, G\"ottingen 37077, Germany}

\leadauthor{I. Pihlajamaa} 

\significancestatement{
Characterising the microstructure of dense liquids remains one of the most challenging open problems in statistical physics. Virtually all research in the last decades focuses on the correlations of the positions between two particles. In this work we go beyond the state of the art and find that the three- and four-body structure reveal many features that two-body one does not, especially at high densities over length scales of a few particle diameters. Moreover, we derive useful approximations for these complex correlation functions involving up to six particles, which are in good agreement with our measurements. These pave the way for the construction of more accurate theories of liquids and their microscopic dynamics.}

\authorcontributions{I.P, C.C.L.L, C.L. and L.M.C.J. designed the research; I.P, C.C.L.L performed the research; I.P, C.C.L.L, C.L. and L.M.C.J. wrote the paper.}
\authordeclaration{The authors have no competing interest to declare.}
\equalauthors{\textsuperscript{1}I.P. contributed equally to this work with C.C.L.L.}
\correspondingauthor{\textsuperscript{2}To whom correspondence should be addressed. E-mail: l.m.c.janssen@tue.nl}

\keywords{Dense liquids $|$ Structural correlations $|$ Density functional theory $|$ Computer simulations $|$ Colloidal hard spheres} 

\begin{abstract}
The complete quantitative description of the structure of dense and supercooled liquids remains a notoriously difficult problem in statistical physics. Most studies to date focus solely on two-body structural correlations, and only a handful of papers have sought to consider additional three-body correlations. Here, we go beyond the state of the art by extracting many-body static structure factors from molecular dynamics simulations and by deriving accurate approximations up to the six-body structure factor via density functional theory. We find that supercooling manifestly increases four-body correlations, akin to the two- and three-body case. However, at small wave numbers, we observe that the four-point structure of a liquid drastically changes upon supercooling, both qualitatively and quantitatively, which is not the case in two-point structural correlations. This indicates that theories of the structure or dynamics of dense liquids should incorporate many-body correlations beyond the two-particle level to fully capture their intricate behaviour.
\end{abstract}

\dates{This manuscript was compiled on \today}
\doi{\url{www.pnas.org/cgi/doi/10.1073/pnas.XXXXXXXXXX}}

\begin{document}

\maketitle
\thispagestyle{firststyle}
\ifthenelse{\boolean{shortarticle}}{\ifthenelse{\boolean{singlecolumn}}{\abscontentformatted}{\abscontent}}{}

The computation of many-body correlation functions remains a central problem in statistical physics. Without them, providing a complete characterisation of an interacting system is often impossible. Examples include the determination of spatial correlations in liquids \cite{hansen2013theory, rahman1976molecular} and charged plasmas \cite{hansen1973statistical}, with the aim  to fully characterise the probability distribution functions of finding a given set of particles at a given set of positions. Other examples include granular media \cite{van1998spatial}, correlated electron systems \cite{kimball1975short,dornheim2018ab} and semiconductors, where quasi-particle excitations can develop highly complex correlations \cite{chemla2001many}.

In most cases, the standard approach to unravelling spatial correlations involves the construction of hierarchies of equations coupling an $n$-th order probability distribution function to an $(n+1)$-th order one. For instance systems where quantum fluctuations are negligible follow the famed Bogoliubov–Born–Green–Kirkwood–Yvon hierarchy \cite{huang2008statistical, hansen2013theory}. Similarly, the many-body Green's functions of a statistical field theory obey the Martin-Schwinger hierarchy \cite{martin1959theory}. These hierarchies are generally truncated using various approximations such that knowledge of the two-body correlations can be extracted. While this has led to pivotal insight in the behaviour of interacting systems, two-body correlations are not always sufficient to fully characterise the behaviour of such systems. This is especially true in the strongly-correlated regime, and it is therefore important to be able to characterise or at least have working approximation schemes for correlation functions beyond the two-body ones.

Specifically, unravelling the microstructure of dense disordered systems such as glasses and supercooled liquids remains a highly challenging, but also very important fundamental problem in liquid state theory \cite{yang2021determining, hansen2013theory, royall2015strong, zhang2020revealing, roth2010fundamental}. In practice, the structure of these systems can be directly measured 
by scattering experiments in the form of the two-body static structure factor $S^{(2)}(\textbf{k})$, where $\textbf{k}$ is the wave vector at which structural correlations are probed \cite{yarnell1973structure,svensson1980neutron,balucani1995dynamics}. 
Precise knowledge of this function, which is also easily obtained from computer simulations, 
gives access to a vast number of a system's thermodynamic and macroscopic properties \cite{hansen2013theory,boon1991molecular,binder2011glassy}. Because of its prevalence in the experimental literature on the liquid state, the two-body static structure factor also has become one of the main quantities used in theoretical development, not only to characterise the structure of liquids but also to predict their dynamical behaviour \cite{gotze1995mode}. However, from a formal standpoint the computation of the structure factor requires knowledge of the three-body correlation function as expressed in the Born-Green-Yvon equation \cite{taylor1992born} or knowledge of the full form of the excess free energy \cite{hansen2013theory}, both of which pose incredibly difficult problems.

Moreover, a collection of recent results points towards the idea that two-body correlation functions such as $S^{(2)}(\textbf{k})$ might not be sufficient to quantitatively describe the structure and the dynamics of very dense liquids. For instance, the existence of a growing static length scale associated with amorphous order near the glass transition has been identified \cite{hocky2012growing,biroli2013comparison,gutierrez2015static,yaida2016point}. This growing length scale is an inherently multi-body one and hence is not captured in the canonical static structure factor. In addition, a plethora of locally preferred, higher-order structures have been identified in numerous glass-forming materials \cite{royall2015structure,tong2019structural,tanaka2019revealing}. For example, metallic glasses have a tendency to prefer localised icosahedral configurations \cite{pedersen2010geometry,wu2015hidden,cheng2011atomic}. Simpler model glass-formers such as Kob-Andersen mixtures also display short-to-medium ranged ordering, often studied via bond-order parameter expansions \cite{steinhardt1983bond,leocmach2012roles,tong2018revealing, boattini2020autonomously}. The presence of these ordered structures is impossible to extract from simple static structure factor measurements as they average out all local angular dependencies by construction. 
Furthermore, higher-order spatial correlation functions have also revealed preferential ordering of alternating layers with icosahedral and dodecahedral symmetries in Kob-Andersen mixtures \cite{zhang2020revealing}, and preferential angular distributions in hard \cite{lehmkuhler2020slowing} and soft \cite{levashov2020structure} particle systems.
More abstract advanced network clustering methods \cite{ronhovde2011detecting} and community inference techniques \cite{paret2020assessing} also detect short-to-medium-ranged ordering in model glass-formers.


All these results indicate that we should expect many-body correlation functions to display highly complex behaviour as one descends in the supercooled regime. It is therefore not unthinkable that these play a large, yet mostly unstudied role in liquid dynamics near vitrification \cite{lehmkuhler2020slowing, schoenholz2016structural, schoenholz2017relationship}. Most studies on static many-body correlation functions so far have focused on triplet correlations, and various factorisation approximations thereof, in both real space \cite{haymet1981triplet,mcneil1983triplet,bildstein1994triplet,gupta1984representation, tanaka1983molecular} and reciprocal space \cite{sciortino2001debye, jorge2002theory, coslovich2013static,donko2017higher, denton1989high}. 
With the notable exception of the work of Zhang and Kob \cite{zhang2020revealing}, who have studied orientationally averaged four-body correlations in real space, no work on higher order spatial correlation functions is known to us. Having accurate measurements or at least valid approximations for many-body structural correlation functions is essential for a fundamental understanding of the dense liquid state.


Here, we present for the first time the four-body structural correlations of dense simple liquids in reciprocal space using both theory and computer simulations. We numerically extract the many-body static structure factors of simulated hard spheres up to fourth order and compare the results  
with convolution approximations obtained from a density functional theoretic approach \cite{barrat1988equilibrium, hansen2013theory}. This work, which can be generalised to even higher orders, provides an important step forward in the full quantitative description and prediction of liquid structure.


\section*{Theory of liquid structure}

We consider a classical multi-component interacting fluid of $N$ particles at bulk number density $\rho_0$. 
The microscopic density of particle species $\alpha$ at position $\BF{r}$ is denoted by $\rho_{\alpha}(\BF{r})$ and the $n$-body density probability distribution by $\rho^{(n)}_{\alpha_1...\alpha_n}(\BF{r}_1,\ldots,\BF{r}_n)$ \cite{barrat1988equilibrium}. The static $n$-body density correlation functions of interest follow from the generalised Ornstein-Zernike integral equations, which can be derived from classical density functional theory \cite{hansen2013theory}. In a translationally invariant system, these functions are defined as correlations of density fluctuations of species $\alpha$, denoted $\hat{\rho}_{\alpha}(\mathbf{r}) \equiv \rho_{\alpha}(\mathbf{r}) - \rho_0$ :

	\begin{equation}
	\begin{split}
	H^{(n)}_{\alpha_1...\alpha_n}(\mathbf{r}_1,...,\mathbf{r}_n)
	\equiv& \ \langle\hat{\rho}_{\alpha_1}(\mathbf{r}_1) \times ... \times \hat{\rho}_{\alpha_n}(\mathbf{r}_n) \rangle \\
	=& \frac{\delta^n \ln (\Xi)}{\delta \ln(z_{\alpha_1}(\mathbf{r}_1))... \delta \ln(z_{\alpha_n}(\mathbf{r}_n))},
	\label{eq:cumulgen}
	\end{split}
	\end{equation}
where $\left<\ldots\right>$ denotes the ensemble average, $\Xi$ is the grand canonical partition function and $z(\mathbf{r})$ is the local activity. Formally, the grand canonical partition function is the cumulant generating functional for the correlation functions $H^{(n)}_{\alpha_1...\alpha_n}$. We also define the functional inverse to $H^{(n)}_{\alpha_1...\alpha_n}$ above as $K^{(n)}_{\alpha_1...\alpha_n}(\mathbf{r}_1,...,\mathbf{r}_n)$ \cite{barrat1988equilibrium}. The inverse functions $K^{(n)}_{\alpha_1...\alpha_n}$ naturally define the many-body direct correlation functions $c^{(n)}_{\alpha_1...\alpha_n}$ from the excess part of the free-energy functional \cite{hansen2013theory}. Details of this derivation are given in the Supplementary Information (SI).

Since the structure of disordered systems is generally studied using scattering techniques, it is useful to work in Fourier space, where the $n$-body density correlation function is proportional to the $n$-body static structure factor $S^{(n)}$ probed at different wave vectors. More precisely, for an isotropic system we write $H^{(n)}_{\alpha_1...\alpha_n}(\mathbf{k}_1,...,\mathbf{k}_{n-1}) = \rho_0 S^{(n)}_{\alpha_1...\alpha_n}(\mathbf{k}_1,...,\mathbf{k}_{n-1})$ \cite{barrat1988equilibrium} where
$S^{(n)}_{\alpha_1...\alpha_n}(\mathbf{k}_1,...,\mathbf{k}_{n-1}) = N^{-1}\langle \hat{\rho}_{\alpha_1}(\mathbf{k}_1)...\hat{\rho}_{\alpha_{n-1}}(\mathbf{k}_{n-1})\hat{\rho}_{\alpha_n}(\mathbf{k}_{n})\rangle$ is the generalised $n$-body structure factor and $\mathbf{k}_j$ is the $j$th wave vector satisfying $\mathbf{k}_i \neq \mathbf{k}_j$ for all allowed $i,j$. This is a necessary condition for the equivalence of the cumulant $S^{(n)}$ with the canonical average of density fluctuations.
We impose momentum conservation by requiring $\sum_{j=1}^{n}\mathbf{k}_j = \mathbf{0}$, simplifying the notation for $n$-point functions in terms of $(n-1)$ arguments. 

The correlation functions $H^{(n)},\ K^{(n)}$ allow for the derivation of generalised Ornstein-Zernike integral equations which become algebraic equations in reciprocal space. These relations are expressed in terms of $S^{(n)}(\BF{k}_1,\ldots,\BF{k}_{n-1})$ and the many-body direct correlation functions $c^{(n)}(\BF{k}_1,\ldots,\BF{k}_{n-1})$ (see SI for a detailed discussion). For $n=3$, it is relatively straightforward to show that the triplet static structure factor is defined as
	\begin{equation}
	\begin{split}
	S^{(3)}_{\alpha\beta\gamma}(\BF{k}_1,\BF{k}_2) = &S^{(2)}_{\alpha\alpha'}({k}_{1})S^{(2)}_{\beta\beta'}({k}_{2})S^{(2)}_{\gamma\gamma'}(|\BF{k}_{1}+\BF{k}_{2}|)\\
	&\times\left( \frac{\delta_{\alpha'\beta'}\delta_{\alpha'\gamma'}}{x_{\alpha'}^2} + \rho_0^2c^{(3)}_{\alpha'\beta'\gamma'}(\BF{k}_1,\BF{k}_2)\right),
	\end{split}
	\label{S3_multi}
	\end{equation}
in which $k_i = |\textbf{k}_i|$,  and $x_{\alpha}$ is the partial fraction of species $\alpha$. We follow Einstein summation convention, summing over repeated indices. Similarly the four-body static correlation function is given by~\eqref{S4_multi}. 
\begin{align}\label{S4_multi}
  S^{(4)}_{\alpha\beta\gamma\sigma}&(\BF{k}_1,\BF{k}_2,\BF{k}_3) = 
  \\\nonumber  &\left( \frac{\delta_{\alpha'\beta'}\delta_{\alpha'\gamma'}}{x_{\alpha'}^2} + \rho_0^2c^{(3)}_{\alpha'\beta'\gamma'}(\BF{k}_1,\BF{k}_2) \right)
  \\\nonumber &\qquad\qquad\times S^{(2)}_{\alpha\alpha'}(k_1)S^{(3)}_{\beta'\beta\sigma}(\BF{k}_1+\BF{k}_2, \BF{k}_3)S^{(2)}_{\gamma'\gamma}(k_2) \\
\nonumber 
 + & \left( \frac{\delta_{\alpha'\beta'}\delta_{\alpha'\sigma'}}{x_{\alpha'}^2} + \rho_0^2 c^{(3)}_{\alpha'\beta'\sigma'}(\BF{k}_1, \BF{k}_3)\right)
  \\\nonumber &\qquad\qquad\times S^{(2)}_{\alpha\alpha'}(k_1)S^{(3)}_{\beta'\beta\gamma}(\BF{k}_1+\BF{k}_3, \BF{k}_2)S^{(2)}_{\sigma'\sigma}(k_3) \\
\nonumber 
 + &\left( \frac{\delta_{\alpha'\gamma'}\delta_{\alpha'\sigma'}}{x_{\alpha'}^2} + \rho_0^2{c}_{\alpha'\gamma'\sigma'}^{(3)}(\BF{k}_2, \BF{k}_3) \right)
  \\\nonumber &\qquad\qquad\times S^{(3)}_{\alpha\alpha'\beta}(\BF{k}_2+\BF{k}_3, \BF{k}_1)S^{(2)}_{\gamma'\gamma}(k_2)S^{(2)}_{\sigma'\sigma}(k_3) \\
 \nonumber -& \left( \frac{2\delta_{\alpha'\beta'}\delta_{\alpha'\gamma'}\delta_{\alpha'\sigma'}}{x_{\alpha'}^3} - \rho_0^3 {c}^{(4)}_{\alpha'\beta'\gamma'\sigma'}(\BF{k}_1,\BF{k}_2,\BF{k}_3) \right)
  \\\nonumber &\qquad\times S^{(2)}_{\alpha\alpha'}(k_1)S^{(2)}_{\beta'\beta}(|\BF{k}_1+\BF{k}_2+\BF{k}_3|)S^{(2)}_{\gamma'\gamma}(k_2)S^{(2)}_{\sigma'\sigma}(k_3).
 \end{align}

Both~\eqref{S3_multi} and~\eqref{S4_multi} are formally exact results. By dropping all indices, we obtain the single-component versions of these equations. An often invoked approximation for such correlation functions 
is the so-called convolution approximation \cite{jackson1962energy, barrat1987factorization}.
This approximation
is obtained by neglecting all contributions from direct correlation functions ${c}^{(n)}(\BF{k}_1,\ldots,\BF{k}_n)$ beyond the two-point one. This is essentially equivalent to neglecting all true $n$-body structural correlations and retaining only those mediated via two-body correlations. This yields
\begin{equation}\label{eq:S3conv}
        S_{\mathrm{conv}}^{(3)}(\textbf{k}_1, \textbf{k}_2) = S(\textbf{k}_1)S(\textbf{k}_2)S(|\textbf{k}_1 + \textbf{k}_2|),
\end{equation}
and
\begin{align}
       &S_{\mathrm{conv}}^{(4)}(\textbf{k}_1, \textbf{k}_2,\textbf{k}_3) \approx S(k_1)S(k_2)S(k_3)S(|\BF{k}_1+\BF{k}_2+\BF{k}_3|)\nonumber\\
    &\times\left( S(|\textbf{k}_1+\textbf{k}_2|)+S(|\textbf{k}_1+\textbf{k}_3|)+S(|\textbf{k}_2+\textbf{k}_3|)-2 \right)
    \label{eq:S4conv}
\end{align}
for monodisperse systems where, following convention, we omit the superscript for the two-body structure factor and denote $S^{(2)}(k)$ by $S(k)$. Although the convolution approximation for $S^{(3)}$ is usually assumed to be 
reasonable for systems with relatively weak attracting interaction potentials \cite{sciortino2001debye}, a recent mode-coupling theory study has revealed that including ${c}^{(3)}(\BF{k}_1,\BF{k}_2)$ can qualitatively change the glass transition diagram even for simple hard-sphere mixtures \cite{luo2022many}. Moreover, the convolution approximation provides even less accuracy for systems such as silica \cite{sciortino2001debye}.
Indeed, silica is part of a family of network forming glasses \cite{della1992molecular} which tend to have strongly anisotropic and attractive interaction potentials due to coordinated bonding. We expect that the failure of the convolution approximation for silica glasses generalises to other anisotropic glass forming materials, where the three- (and higher)-body contributions to the excess free-energy become important. 

For completeness, we also present in the SI the convolution expressions for the five-body structural correlation function $S^{(5)}(\BF{k}_1,\ldots,\BF{k}_4)$ and the six-body structural correlation function $S^{(6)}(\BF{k}_1,\ldots,\BF{k}_5)$ for single component systems, which contain 26 and 236 terms upon full expansion, respectively. 
While testing their validity is beyond the scope of this study (and beyond the scope of current computational efforts), we believe that they might be of utility for physically motivated factorisations of many-body structure factors in first-principles theories of supercooled liquid dynamics.


\begin{figure*}[t]
    \centering
    \includegraphics[width=0.9\textwidth]{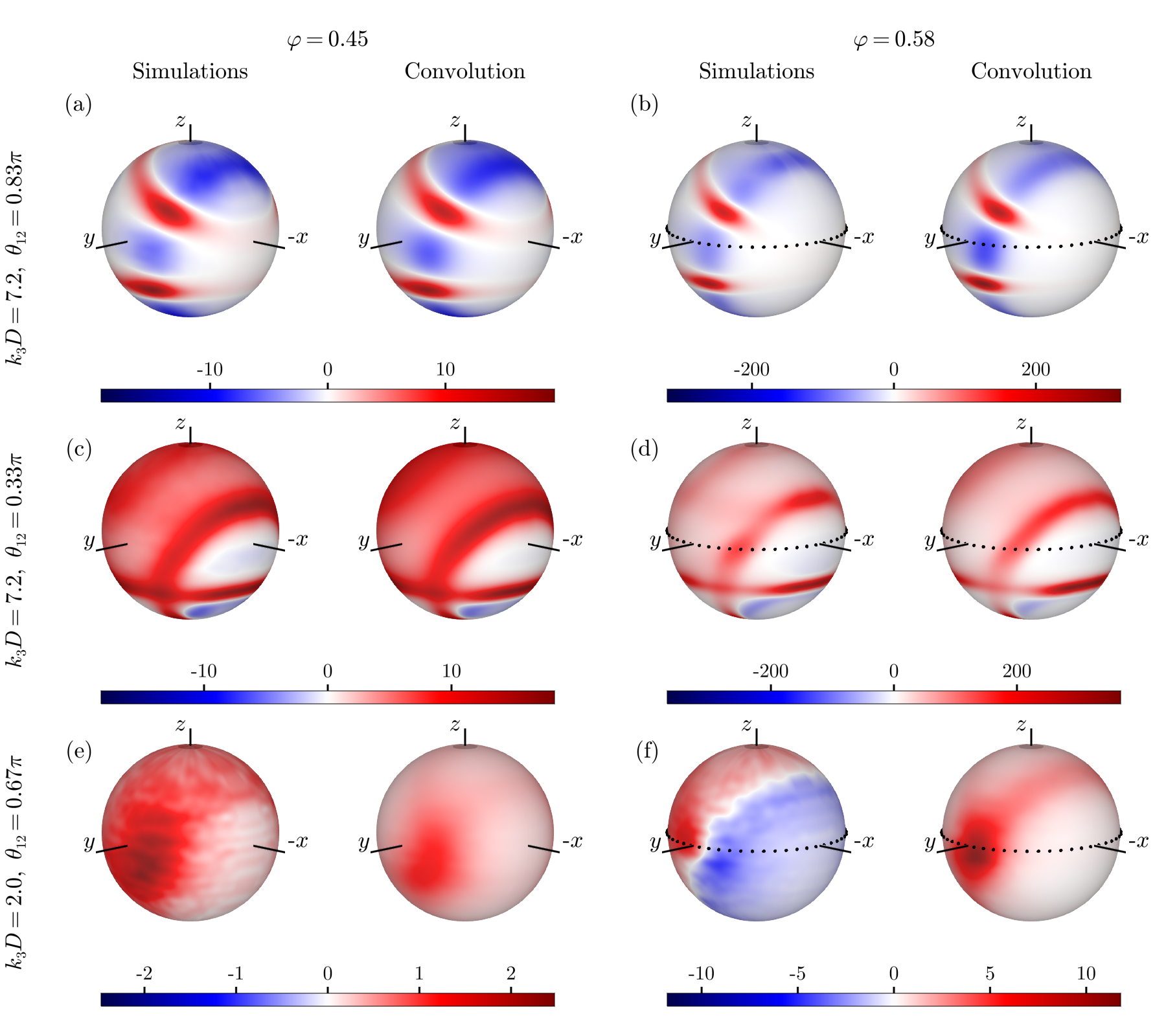}
    \caption{Comparison of simulation results and convolution approximations of the four-point structure factors in normal (left) and supercooled liquids (right). The three rows show $S^{(4)}(k_1, k_2, k_3, \theta_{12}, \theta_{13}, \phi_{23})$ plotted for different values of its arguments, which are specified on the left. In each row, we choose $k_1D = k_2D = 7.2$, corresponding to the main peak of the two-point structure factor. The colours denote the value of the four-point structure factor, which are normalised such that each $S_4$ and its corresponding convolution approximation use the same colour scheme.} 
    \label{fig:S4offdiag}
\end{figure*}

\section*{Comparison with simulations}

To perform a comprehensive test of the convolution approximation for 
$S^{(4)}(\textbf{k}_1, \textbf{k}_2, \textbf{k}_3)$,  we extract the four-body static structure factors directly from numerical simulations. To this end, we perform Monte Carlo simulations of a system of weakly polydisperse hard spheres of averaged diameter $D$, at volume fraction $\varphi$ introduced by Weysser \textit{et al.} \cite{weysser2010structural} (See Materials and Methods). We compare the four-point static structure factor obtained from simulations with its convolution approximation using the following convention.  
The isotropy of our system allows us to rotate the coordinate system such that the $z$-axis coincides with $\textbf{k}_1$, and $\textbf{k}_2$ lies in the $xz$-plane, defining the angle between $\textbf{k}_1$ and $\textbf{k}_2$ as
$\theta_{12}$. The third vector $\textbf{k}_3$ is now determined by the angle $\theta_{13}$ it makes with $\textbf{k}_1$, and the azimuthal angle $\phi_{23}$ which denotes the angle that the projection of $\textbf{k}_3$ on the $xy$-plane makes with that of $\textbf{k}_2$. The latter angle can be expressed as 
\begin{equation}
    \cos \phi_{23} = \frac{k_1^2 (\textbf{k}_2\cdot\textbf{k}_3) - (\textbf{k}_1\cdot\textbf{k}_2)(\textbf{k}_1\cdot\textbf{k}_3)}{\sqrt{k_1^2 k_2^2 - (\textbf{k}_1\cdot\textbf{k}_2)^2}\sqrt{k_1^2 k_3^2 - (\textbf{k}_1\cdot\textbf{k}_3)^2}},
\end{equation}
where $k_i$ is the length of the vector $\textbf{k}_i$. The wave vectors are now given by
  \begin{align}
    \frac{\textbf{k}_1}{k_1} = \begin{pmatrix}
           0 \\
           0 \\
           1
         \end{pmatrix},
         \quad
             \frac{\textbf{k}_2}{k_2} = \begin{pmatrix}
           \sin\theta_{12} \\
           0 \\
           \cos\theta_{12}
         \end{pmatrix},
         \quad
             \frac{\textbf{k}_3}{k_3} = \begin{pmatrix}
           \sin\theta_{13}\cos\phi_{23} \\
          \sin\theta_{13}\sin\phi_{23} \\
           \cos\theta_{13} \\
         \end{pmatrix}
  \end{align}
in Cartesian coordinates.

\begin{figure}[t]
    \centering
    \includegraphics[width=0.44\textwidth]{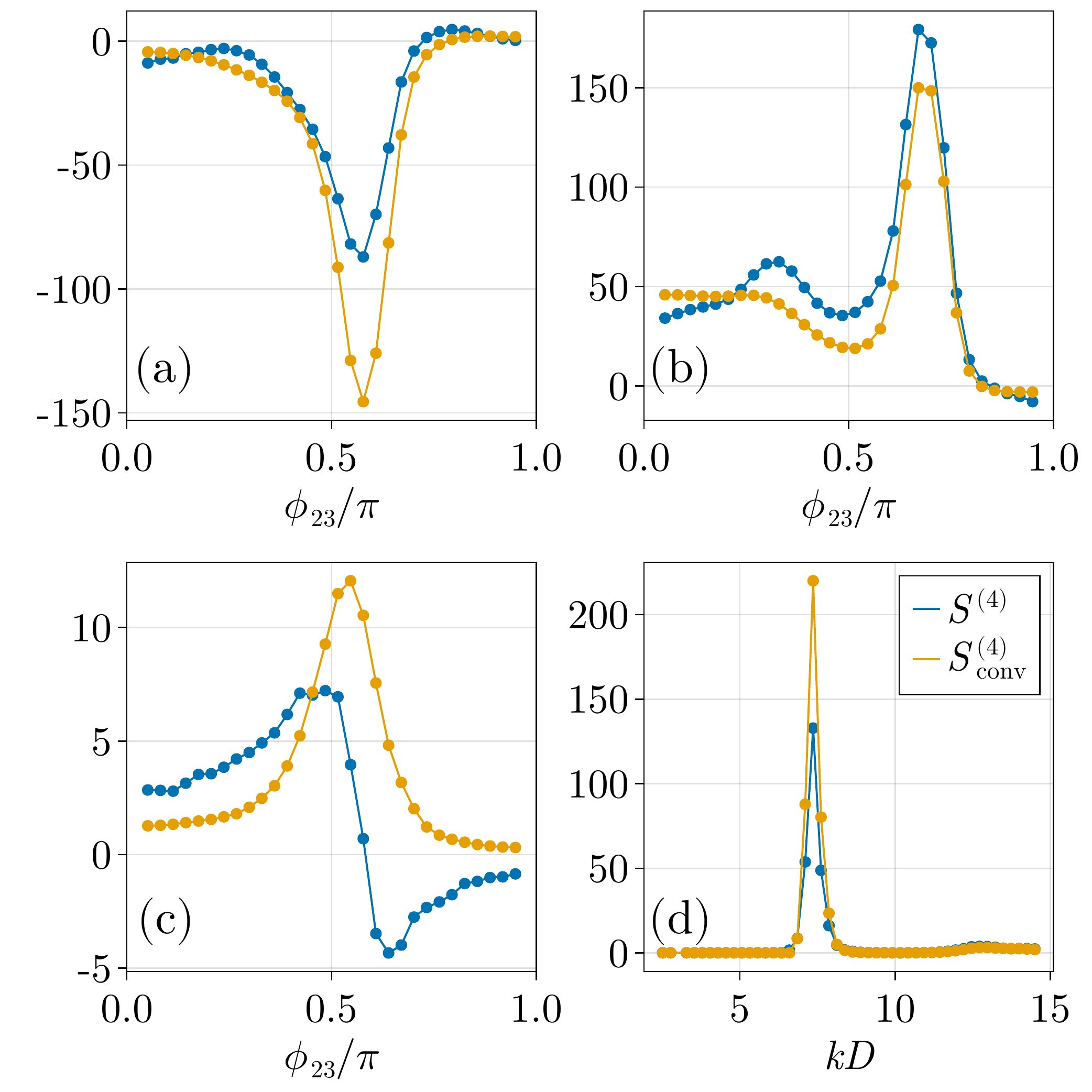}
    \caption{The four-body structure factor and its convolution approximation shown along different contours. Panel (a) shows it along the black dotted contour drawn in Fig.~\ref{fig:S4offdiag}b, panel (b) along that of Fig.~\ref{fig:S4offdiag}d, and panel (c) shows it along the countour in Fig.~\ref{fig:S4offdiag}f. Panel (d) shows the four-point correlator as  a function of $k$ for $k_1=k_2=k_3=k$, $\cos\theta_{12}=1/4$, $\cos\theta_{13}=1/2$, and $\phi_{23}=4\pi/5$.}
    \label{fig:S4lines}
\end{figure}

We show a sample of the results for the four-point structure factor in Figs.~\ref{fig:S4offdiag} and~\ref{fig:S4lines}, in which we show both $S^{(4)}(k_1, k_2, k_3, \theta_{12}, \theta_{13}, \phi_{23})$ measured from simulations and  $S_{\mathrm{conv}}^{(4)}(k_1, k_2, k_3, \theta_{12}, \theta_{13}, \phi_{23})$ obtained from the convolution approximation (\eqref{eq:S4conv}) for low- and high-density liquids at different sets of wave vectors. For purposes of visualisation, we choose to fix the vectors $\textbf{k}_1$ and $\textbf{k}_2$ and the length $k_3$, thereby only varying the angles $\theta_{13}$ and $\phi_{23}$. In this way, the vector $\textbf{k}_3$ traces out the surface of a sphere which we colour according to the corresponding value that $S^{(4)}$ takes. Results for different wave vectors are shown in the SI. In order to make a quantitative comparison, we show in Fig.~\ref{fig:S4lines}(a-c) the same data for the supercooled case plotted along the dotted contours in Figs.~\ref{fig:S4offdiag}. We stress that since we are visualising a function of six scalar variables, it is inevitable that we make arbitrary choices for which wave vectors to analyse. We have inspected the four-body correlations  for many other combinations of wave vectors, which support all our main conclusions.

At intermediate densities in the normal (non-supercooled) liquid regime ($\varphi = 0.45$), we find that the convolution approximation captures both qualitatively and semi-quantitatively the measured four-body correlation function. It manages to reproduce the non-trivial angular dependence, which gives information about the preferred local structure in the liquid \cite{coslovich2013static}. Furthermore, we observe the presence of negative correlations in both $S^{(4)}$ and $S^{(4)}_{\text{conv}}$ depending on the choice of wave vectors. 
We can provide a mathematical reason as $S^{(4)}_{\text{conv}}(\BF{k}_1, \BF{k}_2, \BF{k}_3) < 0$ implies that $\left(S(|\BF{k}_1+\BF{k}_2|) + S(|\BF{k}_1+\BF{k}_3|) + S(|\BF{k}_2+\BF{k}_3|)\right) < 2$ which is for instance satisfied if the wave vectors have similar moduli and their angular separation is large. Indeed, comparing Fig.~\ref{fig:S4offdiag}(a) and (c), we see that near-antiparallel  $\BF{k}_1$ and $\BF{k}_2$ lead to substantially more negative contributions to the four-body correlation function for a fixed $k_3D = 7.2$, since they ensure that $|\textbf{k}_1+\textbf{k}_2|$ and thereby also $S(|\textbf{k}_1+\textbf{k}_2|)$ is small.
Overall it is clear that in normal liquids, the information contained within the two-body structure is sufficient to quantitatively describe many-body structural correlations at least up to the four-body level. 

At higher densities ($\varphi = 0.58$), where the system displays supercooled dynamics, we observe no qualitative changes in the four-body correlation functions for length scales of the order of a particle diameter ($k_3D = 7.2$). We have verified that this also remains true for longer wave lengths, \textit{i.e.} $k_3D > 7.2$ (see SI). However we remark that the correlations already present at low density get amplified by over an order of magnitude at higher density. Previous studies on three- and two-body correlation functions report similar, yet less pronounced, behaviour in the supercooled regime \cite{coslovich2013static, ansell1998structure}. This amplification can also be seen in the functional form of the four-body convolution approximation~\eqref{eq:S4conv}, which scales as the fourth power of the two-body structure factor, while $S^{(3)}$ only scales as its third power. This results in a markedly sharper peak of the four-point structure factor as a function of wave number, shown in Fig.~\ref{fig:S4lines}(d), than is present in the two-point structure factor.

\begin{figure}[t]
    \centering
    \includegraphics[width=0.44\textwidth]{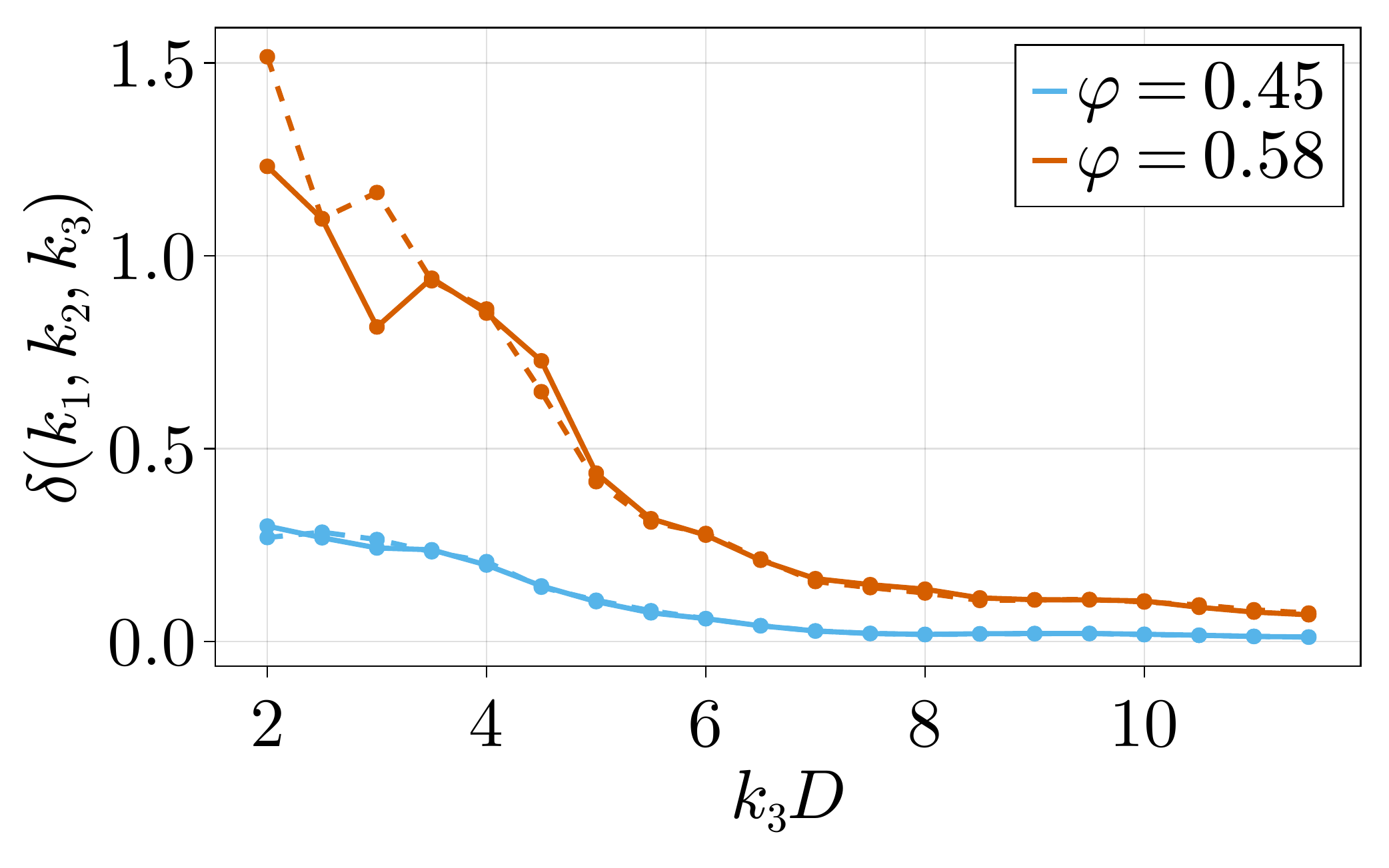}
    \caption{A quantitative measure for the difference between the true four-body structure factor and its convolution approximation at low ($\varphi=0.45$) and high ($\varphi=0.58$) volume fraction. For both volume fractions, the result from two independent simulations are plotted in full and dashed lines in order to show the degree to which statistical noise contributes to this error.}
    \label{fig:delta}
\end{figure}

\begin{figure*}[t]
    \centering
    \includegraphics[width=0.9\textwidth]{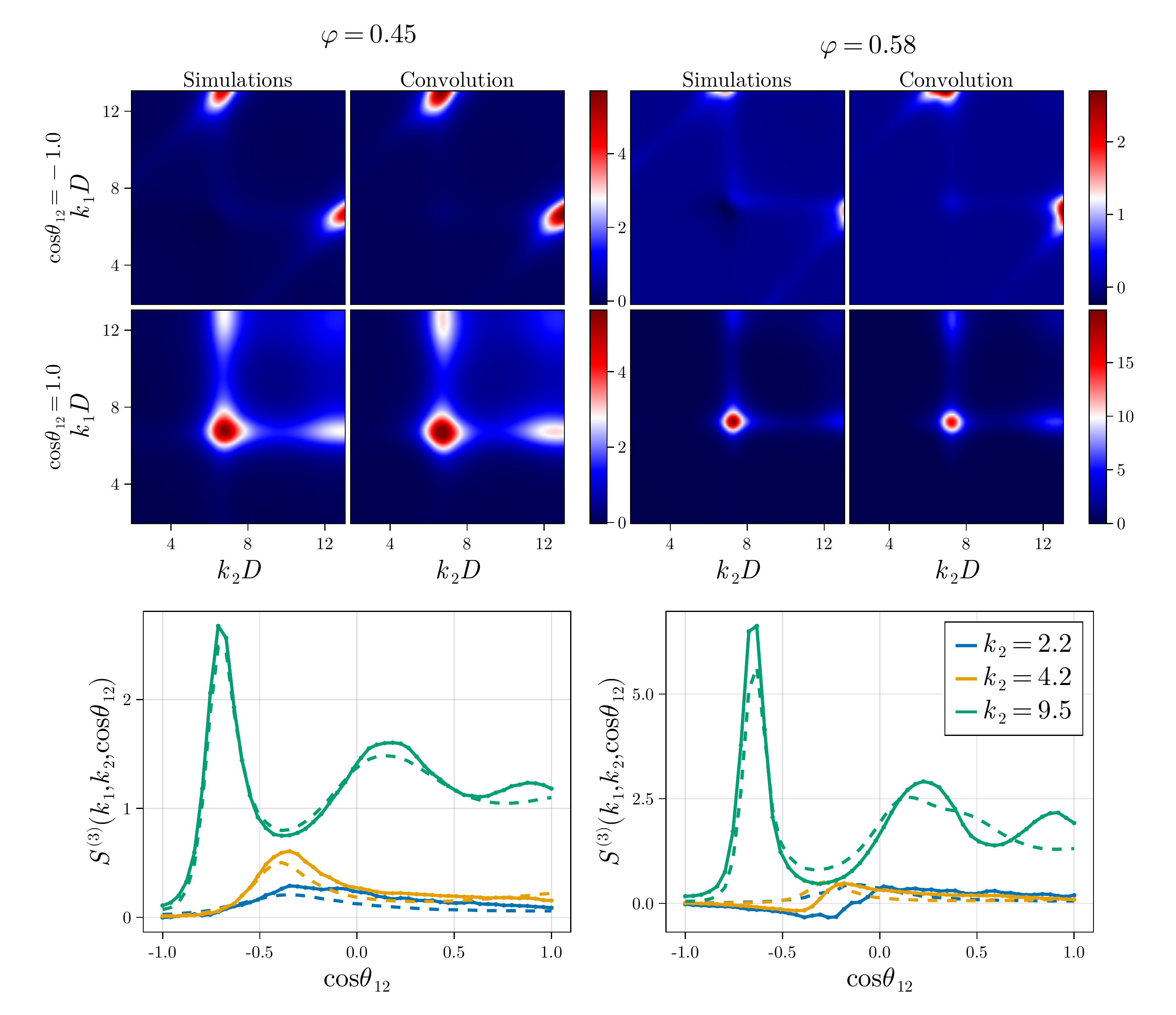}
    \caption{Three-body structure factor compared with its convolution approximation. \textbf{Top:} For low ($\varphi=0.45$) and high ($\varphi=0.58$) densities the triplet correlation function compared with its approximation as a function of the wave numbers $k_1$ and $k_2$ at two different angles $\cos\theta_{12}=\pm1$, corresponding to parallel and anti-parallel configurations. For sake of comparison, the colour scheme is the same in the case of the correlation function and the convolution approximation. \textbf{Bottom:} For low and high densities the triplet correlation function compared with its approximation as a function of the angle between the wave vectors at $k_1D=7.2$ for varying $k_2D$. The dashed lines denote the convolution approximation and the full lines indicate the direct measurements.}
    \label{fig:S3}
\end{figure*}

A strikingly different picture emerges for shorter wave lengths in the supercooled regime. For instance if $k_3D = 2.0$, we see that the four-point structure qualitatively changes with respect to that of a normal liquid, while the convolution approximation does not predict such a change. This means that in supercooled liquids, `true' four-body contributions encapsulated in $c^{(4)}$, are sufficiently dominant in the four-body structure that they qualitatively change the four-body structure at real-space wavelengths of a few particle diameters. This failure of the convolution approximation is also noticeable, albeit less pronounced, for $k_3D = 4.0$ (see SI). We speculate that the marked change of the four-body structure found when supercooling a liquid is caused by the emergence of local structures with some degree of four-fold symmetry, perhaps related to growing four-point dynamic length scales \cite{malins2013identification, lavcevic2003spatially, tanaka2012bond}. 

In order to obtain a quantitative measure of the error of the convolution approximation, which quantifies the degree to which our results are not captured by two-point correlations, we calculate a normalised, angularly averaged difference between the measured $S^{(4)}$ and its convolution approximation. More precisely, we define

    \begin{equation}
        \delta(k_1, k_2, k_3) \equiv \frac{\langle |S^{(4)}(\mathbf{k}_1, \mathbf{k}_2, \mathbf{k}_3) - S^{(4)}_{\text{conv.}}(\mathbf{k}_1, \mathbf{k}_2, \mathbf{k}_3)|^2 \rangle_{\text{ang.}}}{\langle | S^{(4)}(\mathbf{k}_1, \mathbf{k}_2, \mathbf{k}_3) |^2 \rangle_{\text{ang.}}}
    \end{equation}
in which $\langle\ldots \rangle_{\text{ang.}}$ denotes an average over the angles $\cos\theta_{12}$, $\cos\theta_{13}$ and $\phi_{23}$. We show $\delta(k_1,k_2,k_3)$ for fixed $k_1D=k_2D = 7.2$ in Fig.~\ref{fig:delta}. The trend found in the particular cases above seems to be general. Firstly we note that in the low density regime, the error $\delta$ is significantly smaller than in the denser regime considered, corroborating the expectation that the convolution approximation works better in low density cases. Furthermore, we see that at wave numbers smaller the first peak of the structure factor ($k_3D < 7.2$), that is for larger length scales, the error grows significantly. This indicates that on intermediate length scales of a few particle diameters, the convolution approximation fails to correctly capture the microscopic structure. The error $\delta$, as presented in Fig.~\ref{fig:delta}, comprises both the actual error between the four-point correlation function and its convolution approximation as well as the inevitable statistical noise present in our data. In order to show to what extend the latter is present, we have performed the calculation of $\delta$ from the trajectories of two fully independent simulations (full and dashed lines in Fig.~\ref{fig:delta}). We note that the difference between the two lines, and thus the statistical noise in our computation, increases as $k$ decreases. This is caused by the fact that the number of allowed sets of wave vectors $(\textbf{k}_1, \textbf{k}_2, \textbf{k}_3)$ at which we can probe the correlations scales proportional to $k_1^2k_2^2k_3^2$, meaning that we have significantly worse statistics at low $k_3$ than at high $k_3$, for constant $k_1=k_2$. This also causes the visible noise in Fig.~\ref{fig:S4offdiag}(e,f).

To establish to what degree our observations in the four-point structure are also present in the three-body correlations, we conduct a comprehensive analysis of the triplet structure factor $S^{(3)}(\textbf{k}_1, \textbf{k}_2)$ as a function of $k_1$, $k_2$ and $\theta_{12}$. We show a selection of the results in Fig.~\ref{fig:S3}. Indeed, we find a similar phenomenology in the triplet function as we do in the four-body case. That is, at wave numbers around and higher than the first peak of the structure factor, we find that the convolution approximation works well and we see no qualitative changes of the structure upon supercooling. When we probe longer wave lengths, however, both these statements break down. Although less clear than in the four-body case, evidence of a structural transformation can be seen in the lower right panel of Fig.~\ref{fig:S3} for wave numbers below $k_2D \leq 4.2$, with the emergence of a negative dip not observed in the corresponding low density system (lower left panel). This highlights the importance of these high-order density correlations for understanding the supercooled liquid state. Concomitantly, the convolution approximation fails, because it spuriously asserts that all structural information is contained within two-body correlations. Even though we study a slightly different model system, our results qualitatively match those of Coslovich, who reports $S^{(3)}(\textbf{k}_1, \textbf{k}_2)$ and its convolution approximation for the case that $k_1=k_2$ in binary systems \cite{coslovich2013static}. It is plausible that these quantitative changes in the direct triplet correlation function contribute to the qualitative structural transformation we report here since $S^{(4)}$ depends on $c^{(3)}$ (see ~\eqref{S4_multi}). 

\section*{A special case: diagonal four-point correlations}
\begin{figure*}[t]
    \centering
    \includegraphics[width=0.9\textwidth]{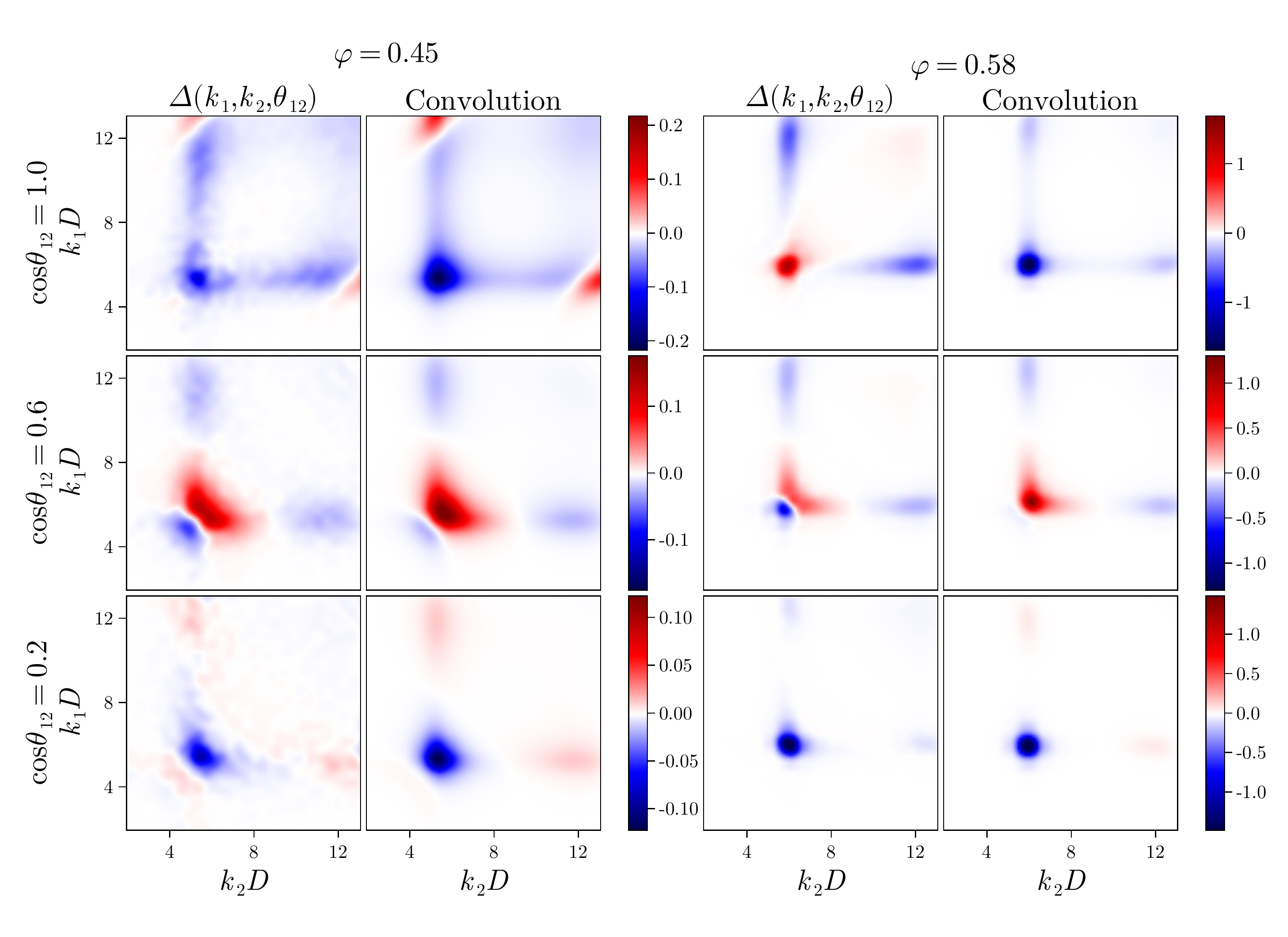}
    \caption{Difference between the diagonal four-point correlation function and the Gaussian factorization approximation $\Delta(k_1, k_2, \theta_{12}) = S^{(4)}(\textbf{k}_1, \textbf{k}_2)-NS^{(2)}(\textbf{k}_1)S^{(2)}(\textbf{k}_2)$ compared with the convolution approximation as a function of $k_1$ and $k_2$ for different values of $\theta_{12}$ in different rows. The left two columns compare simulation results and the convolution approximation for normal liquids, whereas the right compare them for supercooled systems.}
    \label{fig:S4diag}
\end{figure*}

In microscopic theories of liquid dynamics, the four-point structure factor commonly appears in its diagonal form \cite{gotze1995mode, stephen1969raman, halley2012correlation}. This is a special case of the four-body static structure factor which is obtained when the structure is probed at $\textbf{k}_3 = - \textbf{k}_1$ and $\textbf{k}_4 = - \textbf{k}_2$. We refer to the resulting quantity as the \textit{diagonal} four-point structure factor $S_{\mathrm{diag}}^{(4)}(\textbf{k}_1, \textbf{k}_2)$. Note that this is a function only of two independent wave vectors (i.e. two wave numbers and one angle), and therefore may be written as $S^{(4)}_\mathrm{diag}(k_1, k_2,\cos\theta_{12})$. In order to approximate it, the convolution approximation discussed in the above section cannot be applied directly. In fact, we find that the diagonal four-point correlation function very accurately agrees with the so-called Gaussian factorisation approximation $S_{\mathrm{diag}}^{(4)}(\textbf{k}_1, \textbf{k}_2) = NS^{(2)}(k_1)S^{(2)}(k_2) + \mathcal{O}(1)$. Note that within this definition, $S_{\mathrm{diag}}^{(4)}$ scales linearly with the system size, and thus in the thermodynamic limit, the $\mathcal{O}(1)$ term can be neglected. In finite systems, however, this term is measurable and can be approximated by the four-point convolution approximation
\begin{align}\label{eq:S4diagconv}
   & S_{\mathrm{diag}}^{(4)}(\textbf{k}_1, \textbf{k}_2) - NS(k_1)S(k_2) \approx S^{(4)}_\mathrm{conv}(\textbf{k}_1, \textbf{k}_2)\\
    &=S(k_1)^2S(k_2)^2\left(S(0)+S(|\textbf{k}_1+\textbf{k}_2|) + S(| \textbf{k}_1-\textbf{k}_2| )-2\right)\nonumber
\end{align}
as Fig.~\ref{fig:S4diag} shows.

To better understand the nature of these diagonal correlation functions, lets us recall that the many-body structure factors $S^{(n)}$ emerge from the \textit{cumulant generating functional} $\ln (\Xi)$, where $\Xi$ is the grand-canonical partition function. Machta \textit{et al}.~\cite{machta1982mode, halley2012correlation} developed an ordering scheme to identify dominating contributions to cumulant averaged quantities which we use to explain behaviour along diagonals in wave vector space. Essentially, they find that a cumulant average of $n$-linear Fourier transformed densities should scale as $\mathcal{O}(N(\xi/a)^{d(n-1)})$ where $\xi$ is a two-body correlation length, $a$ an average inter-atomic separation and $d$ the spatial dimension. Far from critical points we expect $\xi/a \approx 1$, and it is therefore safe to presume that each cumulant of some product of density modes scales as $\mathcal{O}(N)$. Since there is a formal relation between cumulant averages and standard averages, we can use this information to describe the many-body structure factors. For instance, the cumulant expansion of the pair correlation reads $ \langle \hat{\rho}(\mathbf{k}_1)\hat{\rho}(\mathbf{k}_2) \rangle \delta_{\mathbf{k}_1+\mathbf{k}_2,\mathbf{0}} = \langle\langle \hat{\rho}(\BF{k}_1) \rangle\rangle\langle\langle \hat{\rho}(\BF{k}_2) \rangle\rangle \delta_{\BF{k}_1,\BF{0}}\delta_{\BF{k}_2,\BF{0}} + \langle\langle \hat{\rho}(\BF{k}_1)\hat{\rho}(\BF{k}_2)\rangle\rangle \delta_{\BF{k}_1+\BF{k}_2, \mathbf{0}}$ where we denote cumulant averages with $\langle\langle \ldots \rangle\rangle$ to contrast the standard canonical averages with single angular brackets. The Kronecker deltas are present to enforce translation invariance in these expressions. In this expression, the first term scales as $\mathcal{O}(N^2)$ while the second one scales as $\mathcal{O}(N)$.  If neither wave vector is zero, the canonical average of a pair of density modes coincides with its cumulant average: $\langle \hat{\rho}(\BF{k}_1)\hat{\rho}^*(\BF{k}_1)\rangle  = \langle\langle \hat{\rho}(\BF{k}_1)\hat{\rho}^*(\BF{k}_1)\rangle\rangle$. Similarly, the canonically averaged four-body correlation (which is the quantity that we measure in our computer simulations) can be expanded as
    \begin{equation}
    \begin{split}
    \label{eq:cumexp4}
        &\langle \hrho(\BF{k}_1)\hrho(\BF{k}_2)\hrho(\BF{k}_3)\hrho(\BF{k}_4) \rangle \delta_{\mathbf{k}_1+\mathbf{k}_2+\mathbf{k}_3+\mathbf{k}_4,\mathbf{0}} =\\
        & \langle\langle \hrho(\BF{k}_1) \rangle\rangle \langle\langle \hrho(\BF{k}_2) \rangle\rangle\langle\langle \hrho(\BF{k}_3) \rangle\rangle\langle\langle \hrho(\BF{k}_4) \rangle\rangle \delta_{\BF{k}_1,\BF{0}}\delta_{\BF{k}_2,\BF{0}}\delta_{\BF{k}_3,\BF{0}}\delta_{\BF{k}_4,\BF{0}} \\
        & + \langle\langle \hrho(\BF{k}_1) \hrho(\BF{k}_2) \rangle\rangle \langle\langle \hrho(\BF{k}_3) \rangle\rangle \langle\langle \hrho(\BF{k}_4) \rangle\rangle \delta_{\BF{k}_1+\BF{k}_2,\BF{0}}\delta_{\BF{k}_3,\BF{0}}\delta_{\BF{k}_4,\BF{0}} + \text{ p.} \\
        & + \langle\langle \hrho(\BF{k}_1) \hrho(\BF{k}_2) \hrho(\BF{k}_3)\rangle\rangle\langle\langle \hrho(\BF{k}_4) \rangle\rangle \delta_{\BF{k}_1+\BF{k}_2+\BF{k}_3,\BF{0}}\delta_{\BF{k}_4,\BF{0}} + \text{ p.} \\
        & + \langle\langle \hrho(\BF{k}_1) \hrho(\BF{k}_2)\rangle\rangle\langle\langle \hrho(\BF{k}_3) \hrho(\BF{k}_4)\rangle\rangle \delta_{\BF{k}_1+\BF{k}_2, \BF{0}}\delta_{\BF{k}_3+\BF{k}_4, \BF{0}} + \text{ p.} \\
        & + \langle\langle\hrho(\BF{k}_1)\hrho(\BF{k}_2)\hrho(\BF{k}_3)\hrho(\BF{k}_4) \rangle\rangle \delta_{\BF{k}_1+\BF{k}_2+\BF{k}_3+\BF{k}_4,\BF{0}},
    \end{split}
    \end{equation}
where the terms on each line are of order $\MC{O}(N^4)$, $ \MC{O}(N^3)$,  $\MC{O}(N^2)$, $\MC{O}(N^2)$, and $\MC{O}(N)$ respectively, and all permutations of the wave numbers are denoted as `p.'. It is clear that the dominating terms in this expansion depend on which of the Kronecker deltas survive, which depends on the choice of wave vectors. In the completely off-diagonal contributions, where no subsets of wave vectors sum to the zero vector, only the last term contributes and we recover equivalence between cumulant and canonical averages. However in the diagonal case where $\BF{k}_1 = -\BF{k}_3, \ \BF{k}_2 = -\BF{k}_4$, the dominating term is of order $\MC{O}(N^2)$, with at next-leading order the last term. Hence in this specific case it is more accurate to approximate $S^{(4)}_\mathrm{diag}(\textbf{k}_1,\textbf{k}_2) = N S(k_1)S(k_2) + \MC{O}(1)$, where we have used the fact that $\langle\langle \hrho(\mathbf{k}_1)\hrho^*(\mathbf{k}_1)\rangle\rangle = \langle \hrho(\mathbf{k}_1)\hrho^*(\mathbf{k}_1)\rangle=NS(k_1)$. This is the commonly used Gaussian approximation to the four-point function \cite{gotze1995mode, stephen1969raman}. We can then define 
\begin{equation}
\begin{split}
    \Delta(\mathbf{k}_1,\mathbf{k}_2) &\equiv S^{(4)}_\text{diag}(\mathbf{k}_1,\mathbf{k}_2) - N S(|\BF{k}_1|)S(|\BF{k}_2|),
    \end{split}
\end{equation}
which, according to~\eqref{eq:cumexp4}, can be approximated by the four-point convolution approximation like the fully off-diagonal four-point structure factor.

To verify this we compare $\Delta(\mathbf{k}_1,\mathbf{k}_2)$ and the convolution approximation \eqref{eq:S4conv} in Fig~\ref{fig:S4diag}. We first note that both quantities are symmetric under the transformation $\cos \theta_{12}\to -\cos \theta_{12}$, where $\theta_{12}$ is the angle between $\mathbf{k}_1, \ \mathbf{k}_2$, and therefore we only present results for positive $\cos\theta_{12}$. Similar to the case of the off-diagonal four-point function, we find that $\Delta$ has a very strong angular dependence, both at high and low densities. As expected we see that it is a quantity of order unity, and we have verified that it does not scale with system size. We remark that there is semi-quantitative agreement between the measured and predicted magnitude of the correlations at low densities, but marked qualitative deviations at higher packing fractions, even at wave numbers around the peak of the structure factor. We believe that this discrepancy should be attributed to the neglect of the direct correlation functions of third and fourth order. We stress, however, that in the thermodynamic limit $\Delta$ vanishes in comparison to the Gaussian factorisation, and hence is not needed for a good description of the behaviour of bulk liquids.


\section*{Conclusion}
We have provided the first comprehensive study of four-body structural correlations in reciprocal space for dense liquids. By generalising the two-body static structure factor to higher orders, our work quantifies the structure of disordered systems in terms of two-, three-, and four-body density correlations in the system. We have extracted the many-body structure factors up to fourth order directly from Monte Carlo simulations of dense quasi-hard spheres, and we have derived explicit convolution approximations for them up to sixth order. In principle these efforts may be generalised up to arbitrary order.


For normal liquids, we find that the measured three- and four-point structural correlation functions agree very accurately with the results from the convolution approximations for all wave vectors we studied. Notably, the convolution approximation manages to successfully reproduce the strong angular dependence of the four-body correlation function, which demonstrates that two-body correlations are sufficient to describe the structure of dilute to moderately dense hard-sphere liquids.

In dense (hard-sphere) liquids, however, we do observe qualitative disagreement between the measured three- and four-point structure factors and their convolution approximations beyond length scales of a few particle diameters. This indicates that genuine many-body structural correlations emerge in the dense regime, which may be related to the emergence of locally preferred crystal structures and perhaps to growing four-point dynamic length scales \cite{malins2013identification, lavcevic2003spatially, tanaka2012bond}. These changes in the liquid structure induced upon supercooling might be suggested for use as a probe to distinguish a supercooled state from a liquid one based on structural aspects alone. In future work we intend to link these observations to changes in locally preferred structures of amorphous systems. 

Furthermore, the incorporation of many-body structural correlations is a necessary step in the development of accurate first-principles theories for the dynamics of dense liquids. This work provides appropriate and rigorously derived approximations for these correlations which can be expressed terms of two-body ones only. While we discuss that this is not sufficient for a complete description, expressing many-body correlations in terms of two-body contributions should be preferred over neglecting them altogether \cite{gotze1995mode, hansen2013theory}. In order to go beyond the convolution approximations, frameworks that allow the calculation of the direct correlation functions $c^{(n)}$ could be employed \cite{roth2010fundamental, rosenfeld1989free, rosenfeld1990free}. We speculate that theories describing glassy dynamics need to properly take such many-body correlations into account to improve their flawed predictions in the low-$k$ regime \cite{flenner2013dynamic}.

\matmethods{

We simulate a set of $N=10^3$ particles in a periodic cubic simulation box with volume $L^3$, such that the number density is given by $\rho_0=N/L^3$. In order to approximate hard-sphere behaviour, we let the particles interact according to a strongly repulsive power-law potential
$U_{ij}(r) = \varepsilon k_BT \left(\frac{D_{ij}}{r}\right)^{36},$
where $k_BT=1$ is the thermal energy, $\varepsilon=1/3$  the interaction strength, $r$ the centre-to-centre distance between the particles, and $D_{ij}= [D_i+D_j]/2$ is the average diameter of the particles, in which $D_i$ is the diameter of particle $i$. Particle dispersions interacting with this potential have been extensively studied before, see Refs.~\cite{weysser2010structural, lange2009comparison}, and have been shown to reproduce hard-sphere behaviour. Since monodisperse hard spheres are known to crystallise at high densities, we choose the particle diameters from a uniform distribution $D_i\in(D-\delta, D+\delta)$, where we set the polydispersity parameter to $\delta = 0.1D$. We monitor crystallisation using averaged 4- and 6-fold local order parameters, terminating a simulation run if it displays crystalline structure \cite{steinhardt1983bond,lechner2008accurate}. The degree of crowding in this system can be quantified by a single order parameter for the effective density $\Gamma=D^3\rho\varepsilon^{1/12}$ \cite{lange2009comparison}, which we vary by changing the volume fraction, defined by $\varphi = \pi\rho D^3 (1+\delta^2)/6$, while keeping the interaction strength $\varepsilon$ fixed. To gather statistics, we perform $10^7$ Monte Carlo sweeps, which for the highest volume fraction considered corresponds to roughly $10^2\tau_\alpha$, in which $\tau_\alpha$ is the structural relaxation time of the intermediate scattering function at $kD=7.2$. Every Monte Carlo sweep includes one attempted displacement move for each particle in the system. Every $10^4$ sweeps, we save the particle positions to disk which we later use to compute the many-body structure factors.

The many-body structure factors are most conveniently calculated from their definition in terms of density modes $\hat\rho_\alpha(\textbf{k}, t) = \sum_{j=1}^{N_\alpha} \exp(i \textbf{k}\cdot\textbf{r}_j(t)) - (2\pi)^3\rho \delta(\textbf{k})$. For the purposes of clarity and tractability, we  treat our system as a single-component mixture, thereby neglecting the existence of cross-component correlations. Since the degree of polydispersity is relatively small in our system, we believe that this approximation does not introduce large errors \cite{weysser2010structural}. Hence, we use the monodisperse relations~\eqref{eq:S3conv} and~\eqref{eq:S4conv} for the evaluation of the convolution approximation of the three- and four-body structure factors.

Since we simulate a finite system of particles, there is a fundamental limit on the resolution with which we can choose the $\textbf{k}$-vectors at which we want to probe the density modes. Specifically, the set of allowed $\textbf{k}$-vectors is constrained to $\frac{2\pi}{L} \left[n_x, n_y, n_z\right]$, with $n_x$, $n_y$, and $n_z$ integers. All many-body static structure factors can straightforwardly be calculated from the density modes as tensor contractions; more details are given  in the SI.  In order to properly probe the $n$-body structure factor, we exclude all sets of $n$ wave vectors of which any subset adds to the zero vector, since those cases effectively probe lower order correlations instead, see~\eqref{eq:cumexp4}. To find the convolution approximations, we first obtain the two-point structure factor $S^{(2)}$ from simulations, and subsequently use that to evaluate the convolution approximation~\eqref{eq:S4conv}. The procedure for extracting the three-body and diagonal four-body structure factors is similar. 


}


\showmatmethods{} 

\section*{Data Availability}
The simulation trajectory data, processed data, code to compute triplet and four-point structure factors, and code to reproduce the figures of this work are available at Zenodo with doi: 10.5281/zenodo.7929968.

\acknow{The authors acknowledge financial support from the Dutch Research Council (NWO) through a Vidi grant (IP, CCLL, and LMCJ) and START-UP grant (CL and LMCJ).
}

\showacknow{} 

\bibliography{pnas-sample}

\end{document}